\documentclass[a4paper,10pt,twoside]{cpc-hepnp}
\pdfoutput=1
\usepackage{CJK,upgreek,fancyhdr}
\usepackage{multicol}
\usepackage{graphicx}
\usepackage{booktabs}
\usepackage{amssymb,bm,mathrsfs,bbm,amscd}
\usepackage[tbtags]{amsmath}
\usepackage{lastpage}
\usepackage{graphicx}
\usepackage{epstopdf}
\usepackage{lineno}
\usepackage[colorlinks,citecolor=red]{hyperref}
\usepackage{ulem}
\begin{document}
\fancyfoot[C]{\small 010201-\thepage}


\title{A vertex reconstruction algorithm in the central detector of JUNO}

\author{%
      Qin Liu$^{1}$\email{laoshe07@mail.ustc.edu.cn}%
\quad Miao He$^{2}$
\quad Xuefeng Ding$^{3}$
\quad Weidong Li$^{2}$
\quad Haiping Peng$^{1}$
}
\maketitle

\address{%
$^1$ State Key Laboratory of Particle Detection and Electronics, \\and Department of modern physics,  University of science and technology of China, Hefei 230026,  China\\
$^2$ Institute of High Energy Physics, Chinese Academy of Sciences, Beijing 100049, China\\
$^3$ Gran Sasso Science Insitute(INFN), F. Crispi 7, L¡¯Aquila, AQ 67100, Italy
}

\begin{abstract}
The Jiangmen Underground Neutrino Observatory (JUNO) is designed to study neutrino mass hierarchy and measure three of the neutrino oscillation parameters with high precision using reactor antineutrinos. It is also able to study many other physical phenomena, including supernova neutrinos, solar neutrinos, geo-neutrinos, atmosphere neutrinos, and so forth. The central detector of JUNO contains 20,000~tons of liquid scintillator (LS) and about 18,000 20-inch photomultiplier tubes (PMTs),  which is the largest liquid scintillator one under construction in the world up today. The energy resolution is expected to be 3\%/$\sqrt{E(MeV)}$. To meet the requirements of the experiment, an algorithm of vertex reconstruction, which takes into account time and charge information of PMTs, has been developed by deploying the maximum likelihood method and well understanding the complicated optical processes in the liquid scintillator.
\end{abstract}




\begin{multicols}{2}

\section{Introduction}
 Jiangmen Underground Neutrino Observatory (JUNO) \cite{lab1,lab2,lab3} is an international cooperation experiment which is designed to study neutrino physics. Its main purpose is to determine the mass hierarchy of neutrinos, which is one of the key questions and a fundamental property of neutrino. Meanwhile, a number of other frontier researches associated with neutrino can be carried out, such as the precision measurement of the neutrino mixing matrix elements, the study of supernova neutrinos, solar neutrinos, geo-neutrinos, atmosphere neutrinos and so on.

 JUNO will be constructed in Jiangmen city of GuangDong province, China, located about 700~m underground. It is about 53~km away from the reactors in Yangjiang and Taishan nuclear power plants. The experiment, started construction at the beginning of 2015, is expected to start commission in 2020. The central detector of JUNO is a spherical structure, which contains 20,000 tons liquid scintillator together with about 18,000 20-inch PMTs for photon detection. Besides, there is an acrylic sphere with an inner diameter of 35.4~m, serving as the container of the liquid scintillator. The sphere is immersed into a water pool and supported by a stainless steel truss whose diameter is 40.1~m. The pure water is used to shield radioactive backgrounds from PMTs and surrounding rocks, and about 2000 20-inch PMTs will be installed in the water pool serving as cherenkov detector. The structure of detector is depicted in Fig.~\ref{myfig}, and the detail description of JUNO experiment is in Ref.~\cite{lab1}.
\begin{center}
\includegraphics[width=7cm]{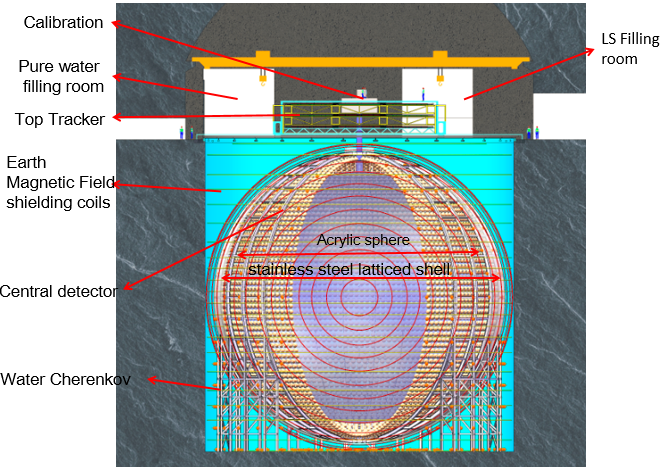}
\figcaption{\label{myfig}   detector structure. }
\end{center}

In JUNO experiment, electron antineutrinos from the reactor are detected via the inverse beta decay (IBD) reaction,
$\overline{\nu}_{e}+p \to e^{+}+n$. The central liquid scintillator detector can measure the oscillated energy spectrum of neutrinos with high precision, i.e., 3\%/$\sqrt{E(MeV)}$.

In order to meet the requirement of energy resolution, the vertex of electron antineutrino must be precisely determined in the first place. Comparing with the running analogous liquid scintillation detectors~\cite{lab4,lab5,lab6,lab7}, the detector volume of JUNO is much larger, which results in the much more significant absorption or scattering for scintillation photons. Additionally, the refractive index is 1.49 for the liquid scintillator while 1.33 for pure water. The large difference of refractive index between two materials results in the large effects of refraction and total reflection at the boundary of two materials, which will affect the time of flight (TOF) in the liquid scintillator for optical photons. Consequently, the optical model is critical for the vertex resolution and is a key factor for JUNO experiment. The performances of the vertex is expected to rely on not only the luminescence time of the liquid scintillator, but also the transit time of the PMT.


The paper is structured as follows. In Section 2, the principle of the vertex reconstruction algorithm is presented. Section 3 gives a detailed study of the determination of TOF and Section 4 describes the probability density function (PDF) and its impact on the vertex reconstruction  performance. Finally, the conclusion is given in Section 5.

\section{Likelihood reconstruction algorithm}
For the neutrinos from the reactors, the energy peaks at 4~MeV, and most of them is less than 10~MeV~\cite{lab8}. The typical spatial track of positrons generated by IBD in the liquid scintillator is a few centimeters, which can be ignored with relative to the detector dimension, 40-meters for diameter. Therefore, the positrons track in the liquid scintillator can be regarded as the point-like light source at the IBD event vertex. And a variable, the residual time
\begin{eqnarray}
\label{myeq1}
t_{i,res} = t_i - {\rm tof}_i - t_0,
\end{eqnarray}
is used to reconstruct the IBD event vertex. Where in Eq.~(\ref{myeq1}),
{\color{red} t$_{i}$ is the first hit time of $i^{th}$ PMT, tof${_i}$ is the corresponding expected time of the photon propagated from the IBD event vertex to the PMT, and t$_0$ represents the occurred time of an IBD event. In the liquid scintillator, when an IBD event occurred, the produced positron deposits all its energy in the scintillator within a negligibly short time and distance. The deposited energy stimulates luminescence of liquid scintillator in a time interval which follows a probability function of summing two exponential functions presenting for the fast and slow components of lights, individually.  Then the emitted photons propagate from the IBD vertex to the photocathode of PMT, converse into electrons, multiplicated and collected by the PMT. The photons propagated time in the scintillator (tof${_i}$) is expected to be proportional to the path length, and the transition time in PMT is a constant for a given PMT, which can be calibrated offline practically and omitted in this study (only its uncertainty will be considered later).  }

Apparently, tof${_i}$ depends on the positions of IBD event vertex and the PMT. Therefore, the vertex $\vec {R}(x,y,z)$ and the IBD event start time t$_0$ are the unknown parameters in Eq.~(\ref{myeq1}), and can be extracted by performing a maximum likelihood fit on the joint likelihood function running over all the PMTs,
\begin{equation}
\label{myeq_likelihood}
  \mathcal{L}=\prod_{\substack{i}}
  f(t_{i,res}),
\end{equation}
 where $f(t_{i,res})$ is the probability density function (PDF) for the residual time of the $i^{th}$ PMT. PDF $f(t_{i,res})$ is the sum of two components,  $i.e.$, a double exponential function presenting for the luminescence time of liquid scintillator, and a Gaussian function taking into account the uncertainties associated with PMT transition time and that contributed from the trajectory of lighting including the effects of muti-scattering, Reyleigh scattering and re-emission $etc$. The studies of time of flight and PDF will be presented in detail in Sec.\ref{sec.tof} and Sec.\ref{sec.pdf}, respectively.

The result of likelihood fitting approach is usually sensitive to the initial inputs of the parameters to be fitted. The initial event start time is set as $t_{0}$ and the initial vertex is determined by the charge-weighted approach which is a typical way to estimate the event vertex roughly. Assuming there is a point source in the liquid scintillator which emits photons isotropically, the closer the PMT to the source, the more photons will be detected. By considering the collected charge as weight of each PMT position, the event vertex can be estimated through the following formula:
 \begin{equation}\label{myeqQ}
 \vec {r_0}=\frac{\sum_{i}q_{i}\vec {r_i}}{\sum_{i}q_{i}},
 \end{equation}
where $\vec r_{i}$ is the position of the $i^{th}$ PMT and $q_i$ is the readout charge of $i^{th}$ PMT.
Suppose the true event vertex is z$_{0}$, one can calculate the vertex derived from the charge-weighted approach:
\begin{eqnarray}\label{myeqz}
  <z> &=& \frac{1}{4\pi}\int zd\Omega\nonumber\\
    &=& \frac{1}{4\pi}\int_0^{2\pi}d\phi\int_0^{\pi}(z_{0}+r\cdot\cos\theta)\sin\theta d\theta\nonumber\\
    &=& \frac{1}{2}\int_0^{\pi}-(z_{0}+(\sqrt{R^2-z_0^2\sin^{2}\theta}-z_0\cos\theta)\cdot\cos\theta) d\cos\theta\nonumber\\
    &=& \frac{1}{2}\int_{-1}^{1}(z_{0}+x\sqrt{R^2-z_{0}^{2}x^{2}}-z_{0}x^2)dx\nonumber\\
    &=& \frac{2}{3}z_0\nonumber.
\end{eqnarray}

{\color{red}Once the reconstructed vertex z is obtained, a better approximation of $z_0$ can be got by multiplying z by a factor of $3/2(1.5)$. The factor 1.5 is derived from mathematics purely, regardless of the geometry of detector and any optical processes in experiment.  GEANT4\cite{lab9} Monte Carlo (MC) simulation shows a factor of 1.2 after considering detector response, in particular, the attenuation of lights in the liquid scintillator and the optical coverage of PMTs. As a result, 1.2 is chosen in the algorithm instead of 1.5.}

\section{Time of flight in the detector}\label{sec.tof}

Time of flight of a scintillation photon starts from the time it is generated to that it is detected by a PMT. In principle, all the hit time in PMT should be used to reconstruct vertex. However, if the time intervals of two or more scintillation photons hitting on the same PMT are too small, they can not be distinguished, especially when the deposit energy is large or the event vertex is close to the edge of the detector. Thus, the most conservative way is to use the first hit time only.

Ideally, the optical path of a photon is considered as a straight line between event vertex and PMT. It is the most common situation for the detectors whose size is not so large. However, JUNO should be studied carefully because of its large size. In addition, there is a big difference between the refractive index of liquid scintillator and that of water. This may result in more complicated optical process, i.e., total reflection.

As is known to us all, many optical processes, such as refraction, absorption, re-emission and Rayleigh scattering~\cite{lab9}, will change the light propagation. The larger the size of a detector, the more significant the effect will be. It turns out that the actual optical path of a scintillation photon is longer than the length of straight line assumed and the expected TOF is smaller than the actual one.

\subsection{Effective velocity}
Due to the fact that scintillation photons are generated with different wavelengths, resulting in different refractive indices and velocities, the group velocity should be used to predict the light propagation instead of the phase velocity. According to Ref.~\cite{lab10}, the dispersion can be parameterized by the Sellmeier equation
\begin{equation}\label{101}
  n^2(\lambda) = 1 + \frac{B}{1-C/\lambda^2},
\end{equation}
where B and C are fitting parameters. The phase velocity is given by the following formula
\begin{equation}\label{102}
  V_p = \frac{\omega}{k} = \frac{c}{n},
\end{equation}
 where $c$ is the light speed in vacuum and n is the refractive index of liquid scintillator.
We can calculate the group velocity according to $V_g=\mathrm{d}\omega/\mathrm{d}k$, given $\lambda = 2\pi c/\omega$ and the Sellmeier equation, we can obtain the following expression:
 \begin{equation}\label{103}
   V_g = \left(\frac{c}{n}\right)\left(1-\frac{\lambda}{n}\frac{\mathrm{d}n}{\mathrm{d}\lambda}\right)^{-1} = V_p\left(1-\frac{\lambda}{n}\frac{\mathrm{d}n}{\mathrm{d}\lambda}\right)^{-1}.
 \end{equation}

\begin{center}
\includegraphics[width=7cm]{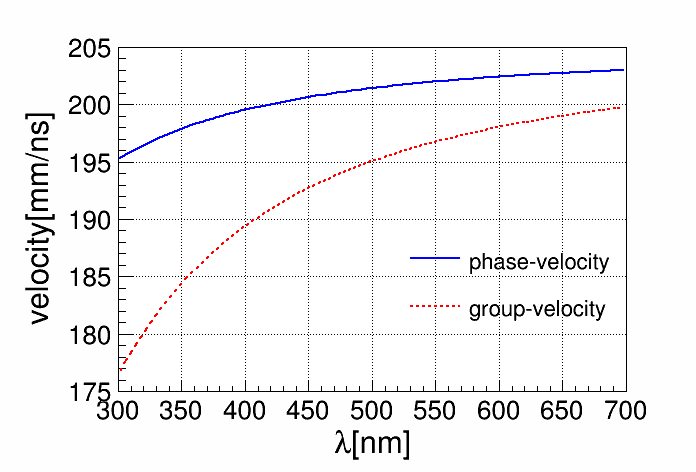}
\figcaption{\label{myfig202}   phase velocity/group velocity v.s. wavelength.}
\end{center}

Fig.~\ref{myfig202} shows the velocity results. Combined with the luminescence spectrum of liquid scintillator, the mean phase velocity and the mean group velocity are $200.5 \pm 3.0$~mm/ns  and $192.2 \pm 3.0$~mm/ns, respectively.

In fact, the "effective velocity" is used in the vertex reconstruction algorithm and it is obtained from the Monte Carlo simulation. The effective velocity can be obtained by the following steps: 1. Set LS luminescence time to 0ns; 2. Put $\gamma$ source at different places (0,0,0~m), (0,0,1~m), (0,0,2~m), ..., (0,0,x~m) 3. Get the first hit time distribution for a certain PMT, e.g., @(0,0,19.5~m); 4. Calculate the effective velocity according to $v_{eff}=l/t$, where l is the distance from source to PMT and t is the peak time of the distribution. The MC simulation shows that the effective velocity is 194.8~mm/ns and seldom changes with different $\gamma$ source places, which is very close to the mean group velocity. The difference between the mean group velocity and the simulation result may come from the water buffer area and the configuration of the detector. By deploying the effective velocity, the reconstructed vertex results are improved significantly.

\subsection{Absorption and re-emission}
Scintillation photons might be absorbed by solvent or solute of liquid scintillator during propagation and then re-emitted. As a result, the wavelength of photons will change. During this process, short wave-length photon is absorbed and long wave-length photon is emitted isotropically.

The place where re-emission process takes place is close to the event vertex within about several tens of centimeters. The relationship between effective refractive index and optical path of a photon is studied by MC simulation and it shows that the effective velocity changes less than 1\% as optical path changes, which can be neglected safely.

\subsection{Refraction}
To revise TOF, refraction and total reflection between liquid scintillation and water are considered.
Fig.~\ref{myfig3} demonstrates the optical model.
\begin{center}
\includegraphics[width=7cm]{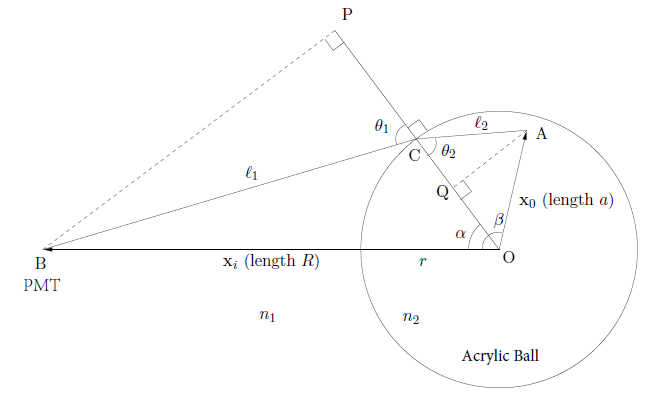}
\figcaption{\label{myfig3}   Optical model used to calculate the optical path of a photon from a given vertex to a specific PMT. A is the event vertex, O is the center of the detector, B is PMT's photocathode and C is the point where photon is refracted.}
\end{center}
Then, equations below are established through geometry relationship and Snell's law.
\begin{eqnarray}
\left\{
\begin{array}{lll}
  l_1\cos\theta_1 &=& R\cos\alpha -r \nonumber\\
  l_1\sin\theta_1 &=& R\sin\alpha \nonumber\\
  l_2\cos\theta_2 &=& r - a\cos(\beta - \alpha) \nonumber\\
  l_2\sin\theta_2 &=& a\sin(\beta - \alpha) \nonumber\\
  n_1\sin\theta_1 &=& n_2\sin\theta_2 \nonumber
\end{array}
\right.
\end{eqnarray}
The relationship between $\alpha$ and $\beta$ is
\begin{equation}\label{myeqrelation}
\begin{aligned}
  1+\left[\frac{r}{a}\csc(\beta-\alpha)-\cot(\beta-\alpha)\right]^2= \\
  \left(\frac{n_2}{n_1}\right)^2\left[1+\left(\cot\alpha-\frac{r}{R}\csc\alpha\right)^2\right]
\end{aligned}
\end{equation}
Once $\alpha$ is determined, TOF can be calculated through
\begin{eqnarray}
\left\{
\begin{array}{lll}
  l_1^2 &=& R^2+r^2-2Rr\cos\alpha \nonumber\\
  l_2^2 &=& a^2+r^2-2ar\cos(\beta-\alpha) \nonumber\\
  TOF &=& (n_1l_1+n_2l_2)/c.\nonumber
\end{array}
\right.
\end{eqnarray}


\begin{center}
\includegraphics[width=7cm]{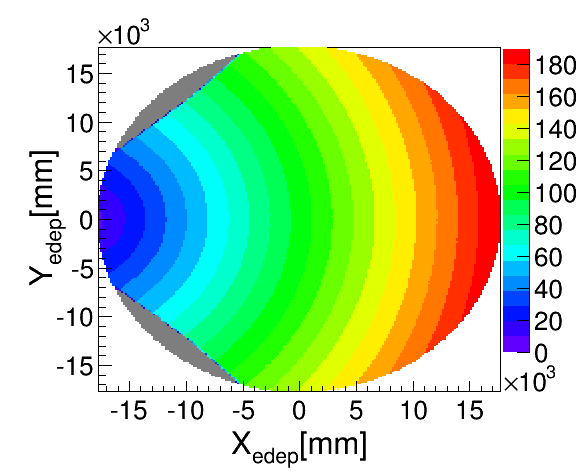}
\figcaption{\label{myfig5}   TOF distribution on plane XY. The time of flight function from a test point to a PMT located at (-R,0,0). The TOF is shown on XY-plane and the grey regions represent "dark zone". Unless it is scattered, a photon cannot travel from a point in the dark zone of a PMT to the center of that PMT's photocathode.}
\end{center}

Fig.~\ref{myfig5} shows the TOF distribution on plane XY. There is an obvious ¡°dark zone¡± (grey area) due to total reflection. Fig.~\ref{myfigtofcmp} shows the difference between straight line case and refraction case. The result demonstrates that TOF are almost the same when vertex is close to center of detector. However, TOF split into two parts when vertex is close to the edge of the detector. They correspond to near-end PMTs and far-end PMTs separately. TOF can be 8ns greater than that of the straight line situation as the optical path of a scintillation photon becomes large.
\begin{center}
\includegraphics[width=7cm]{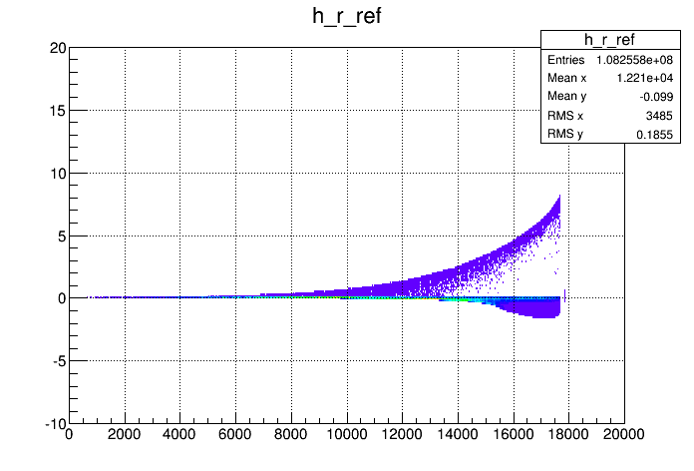}
\figcaption{\label{myfigtofcmp}   TOF comparison. }
\end{center}


\subsection{Total reflection}
The refractive index of liquid scintillator and water are 1.49 and 1.33 respectively. So, according to the following formula:
\begin{eqnarray}
\label{myeq4}
r_c = R_{LS}\times \frac{n_{water}}{n_{LS}},
\end{eqnarray}
the total reflection takes place only when the distance between event vertex and detector center is greater than 16~m.
Theoretically, photons in the dark zone of a certain PMT can not be detected by that PMT. In reality, however, there are still photons that can travel from a point in the dark zone of a PMT to the center of that PMT's photocathode due to Rayleigh scattering and edge effect.

\begin{center}
\includegraphics[width=7cm]{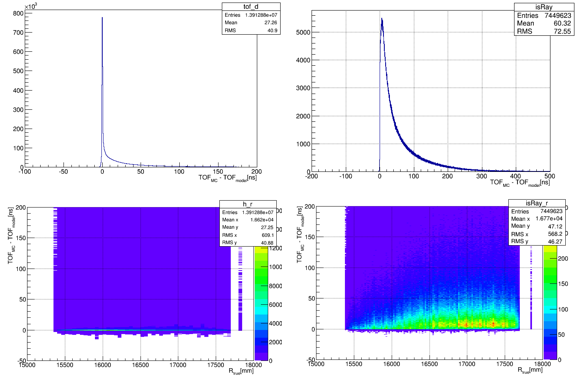}
\figcaption{\label{darkzone}   TOF in dark zone comparison between MC and model. Most estimations are consistent with simulation. For Rayleigh scattering photons(second column), the bias becomes large and is broaden widely. The closer to the edge, the worsen the estimation is. }
\end{center}

Fig.~\ref{darkzone} shows
the difference of TOF between MC simulation and expected.
The true TOF of photon which Rayleigh scattering occurs is about 10~ns greater than expected.
To sum up, in the dark zone of a PMT, direct photons can be detected by that PMT, as well as indirect photons which Rayleigh scattering takes place. For direct photons, optical model can describe their behaviors well, but the model for scattering photons still need to be further studied. We have tried to remove signals of dark zone, but the result even becomes worse.


\subsection{Rayleigh scattering}
Rayleigh scattering can be thought as elastic scattering which can only change the direction of propagation of photons. It will spread the time of flight of PE. The scattered photons will follow the well-known angular distribution predicted by Rayleigh's theory\cite{lab11}, which can be characterized by the volume scattering function $\beta(\theta)$,
\begin{equation}\label{myeqray}
\beta(\theta)=R\left( 1+\frac{1-\delta}{1+\delta}\cos^2\theta \right),
\end{equation}
where $\delta$ is the depolarization ratio and $R$ is volume scattering function when the scattering angle is $90^{\circ}$, i.e., $R\equiv\beta(90^{\circ})$.

However, it is pretty challenging to predict the time of flight of each PE. Currently, the strategy is set a time window and abandon signals outside the time window. Fig.~\ref{myfigpdf} shows that the PDF value is mainly distributed in the area from t$_{res}$ = -5~ns to t$_{res}$ = 30~ns which is the range of the time window. By this way, we can reduce the impact of Rayleigh scattering.




\section{Probability density function}\label{sec.pdf}

In the liquid scintillator detector, the PDF of the residual time of single photon can be described as:
 \begin{eqnarray}
 \label{myeq2}
 f_{res}(t) &=& \frac{1}{ \sqrt{2\pi}\sigma}\exp \left\{ -\frac{(t-t_{0})^{2}}{2\sigma^{2}} \right\}\nonumber\\
 &&\otimes\left[\frac{\omega}{\tau_1}e^{\frac{t}{\tau_{1}}}+\frac{1-\omega}{\tau_2}e^{\frac{t}{\tau_{2}}}\right].
 \end{eqnarray}
 It is the convolution of two terms. The exponential term represents the luminescence time of the liquid scintillator where $\tau_1$ and $\tau_2$ correspond to the fast and slow components. The gaussian term is responsible for the systematic uncertainty of the residual time, which is dominated by  the transit time spread (TTS) of PMT. 



If a PMT detects N photons, the PDF of the residual time of the first hit f(t,N) can be derived from Eq.~\ref{myeq3}
 \begin{equation}
 \label{myeq3}
 f(t,N) = Nf(t)\left(\int_t^{+\infty}f(x)dx\right)^{N-1},
 \end{equation}
 where f(t) corresponds to the PDF of single photon.
 PDFs of different number of photons are shown in Fig.~\ref{myfigpdf}. Apparently, the more photons are detected, the earlier the first hit time is, and the higher the time resolution is.

 \begin{center}
\includegraphics[width=7cm]{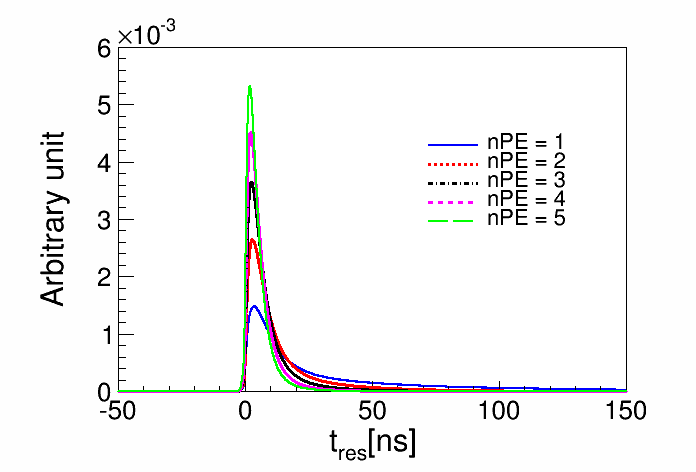}
\figcaption{\label{myfigpdf}   PDF for different number of photons. The more photons detected, the sharper the PDF is. When the number of photons is greater than 5, the PDF are similar and obey the gaussian distribution.}
\end{center}

\subsection{Number of PDF}
By default, 5 PDFs are deployed in reconstruction algorithm. We define a ratio: the number of PMTs that detect photons no more than 5 to the number of all PMTs that detect photons. From Fig.~\ref{myfig6}, we can see the relationship among ratio, deposit energy and event vertex. As we can see that when energy becomes greater or vertex get closer to the edge of detector, the ratio turns to smaller which means more number of PDF should be considered. In algorithm, if a PMT detects photons greater than 5, we still deploy 5-PE PDF because simulation shows that the PDF is similar when number of PE is greater than 5.
\begin{center}
\includegraphics[width=7cm]{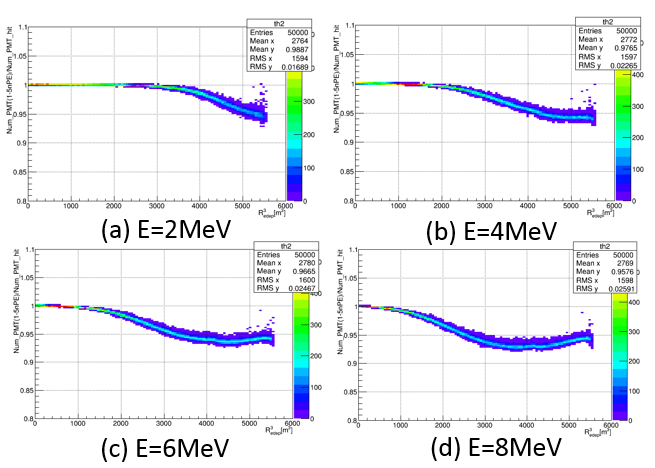}
\figcaption{\label{myfig6}   Relationship among ratio, the number of PMTs that detect photons no more than 5 to the number of all PMTs that detect photons, position and energy of an event. }
\end{center}

\subsection{Difference between $\gamma$ and $e^+$ source}
Experimentally, the residual time distribution can be accessed by a particle source at the detector center. Traditional calibration source is $\gamma$ source. However, the space dispersion of a $\gamma$ ray is relatively large, therefore it can influence the resulted residual time distribution. For comparison, we show the calibrated residual time distribution of two different sources from MC simulation in Fig.~\ref{myfig7}.

Positron deposits energy in liquid scintillator by two steps: first, it deposits its energy until its kinetic energy becomes zero, then annihilates with an electron and emits a pair of gamma whose energy is 0.511~MeV. For low energy $e^{+}$ events, the source is similar to a $\gamma$ source. For high energy $e^{+}$ events, using $\gamma$ source PDF may introduce a little bias of vertex reconstruction results. If possible, we may choose a proper source to obtain PDF. However, for now, positron source is not accessible in calibration.
\begin{center}
\includegraphics[width=7cm]{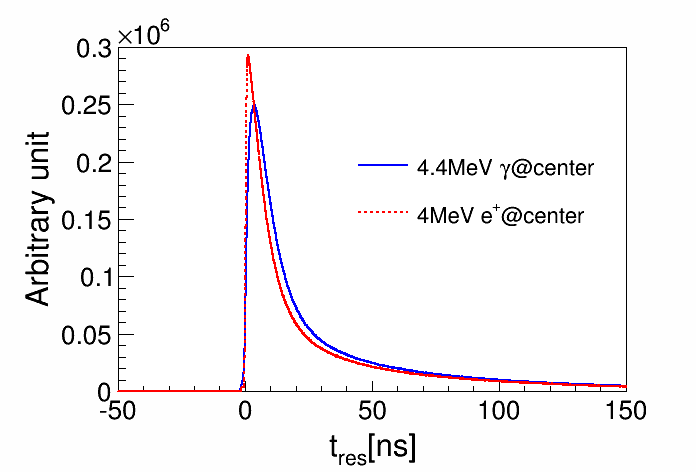}
\figcaption{\label{myfig7}   PDF comparison between 4.4~MeV $\gamma$ source and 4~MeV $e^{+}$ source at detector center. Due to the fact that space dispersion effect of $\gamma$ is much larger than that of $e^+$, the PDF shape of $e^+$ is much sharper than that of $\gamma$.} 
\end{center}


\subsection{PMT TTS impact on vertex reconstruction}
Fig.~\ref{myfigvbias} and Fig.~\ref{myfigvsigma} illustrate the performance of vertex reconstruction. Bias of vertex is less than 3~cm in fiducial volume(R $\le$ 17.2~m). 
\begin{center}
\includegraphics[width=7cm]{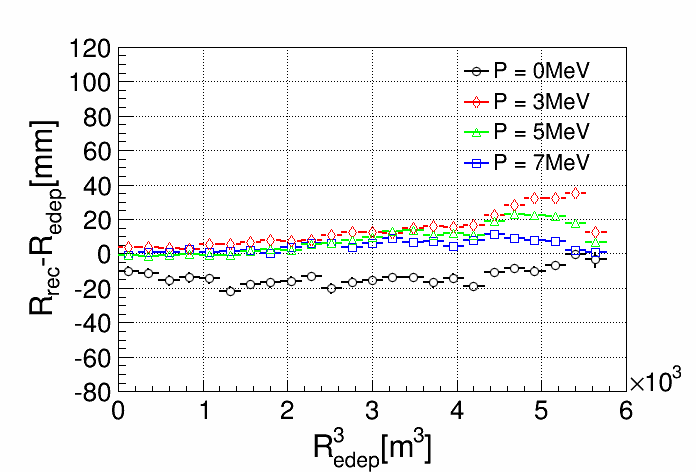}
\figcaption{\label{myfigvbias}  Reconstructed vertex bias. By deploying the 5-PDFs, considering refraction and set a time window from -5~ns to 30~ns, the bias of reconstructed vertex is less than 4~cm without any correction.}
\end{center}
\begin{center}
\includegraphics[width=8cm]{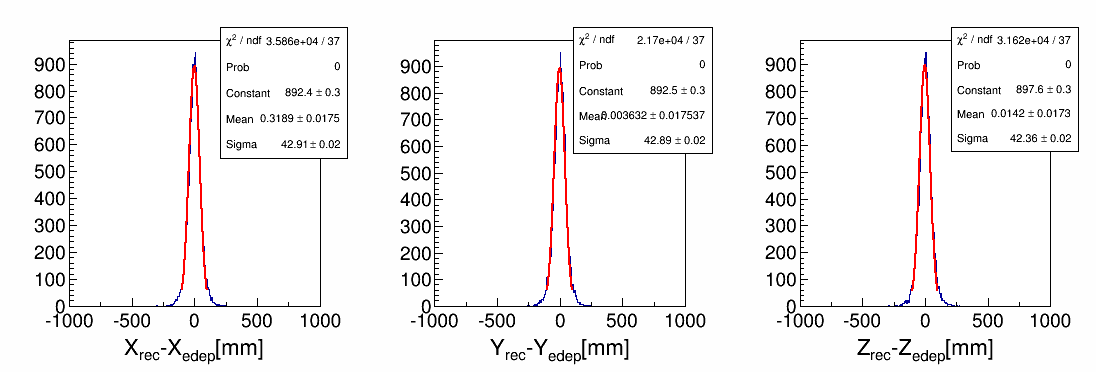}
\figcaption{\label{myfigvsigma}  Reconstructed vertex resolution of X, Y, Z direction. The mean values are consistent with MC truth and $\sigma$ are similar for three directions.}
\end{center}

Currently, the TTS of 20 inch PMT can be as large as 20~ns\cite{lab12}. i.e., $\sigma$ of time resolution of PMT will be 8~ns, according to the relationship TTS = 2.354$\sigma$. Compared with liquid scintillator luminescence time, the fast component of which is 4.93~ns and the slow component is 20.6~ns, PMT TTS will dominate the uncertainty of residual time, especially when vertex is close to PMT. Fig.~\ref{myfig9} shows how PMT TTS influences vertex resolution. If TTS is not considered, the vertex resolution is about 7~cm@1~MeV. If TTS is about 10~ns, vertex resolution increases to 11~cm@1~MeV. In the future, 3 inch small PMTs will be added. Its TTS can be as good as 1ns which is much better than 20 inch PMT. The vertex resolution could improve, but how good it can reach is beyond the content of this article and needs further study.
\begin{center}
\includegraphics[width=7cm]{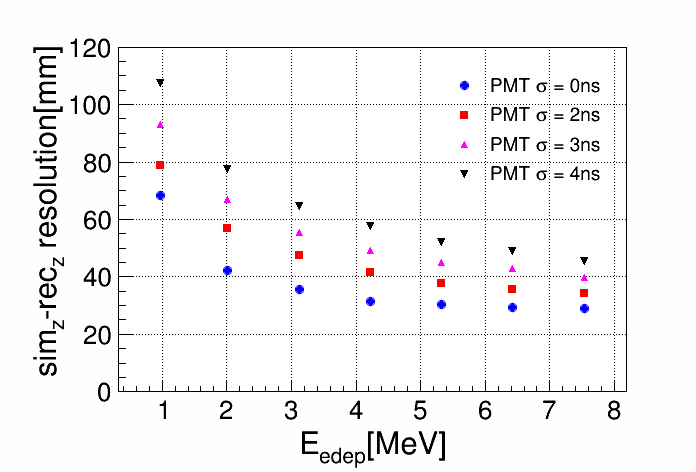}
\figcaption{\label{myfig9}   PMT TTS impact on vertex resolution.  }
\end{center}

\section{Conclusion}
 Jiangmen underground neutrino observatory (JUNO) is a multi-purpose neutrino experiment which is under construction. Its large size and materials used in the central detector makes it unique compared with other running detectors. To meet the requirement that the energy resolution is expected to be 3\%/$\sqrt{E(MeV)}$, an algorithm of vertex reconstruction has been developed. This algorithm use the maximum likelihood method by taking account of time and charge information of PMTs, combining the study of optical model in the detector. Preliminary results of vertex reconstruction has been introduced. By deploying reconstruction algorithm, the reconstructed vertex bias is less than 3~cm in fiducial volume and vertex resolution is about 7~cm@1~MeV if TTS is not considered.

\section*{Acknowledgements}

This work is supported by National Science Foundation for Distinguished Young Scholars of China (Grant No. 11625523),
National Natural Science Foundation of China (Grant No. 11575226, 11575224), and the Strategic Priority Research Program of the Chinese Academy of Sciences (Grant No. XDA10010900).

\end{multicols}

\vspace{15mm}

\vspace{-1mm}
\centerline{\rule{80mm}{0.1pt}}
\vspace{2mm}

\begin{multicols}{2}

\end{multicols}

\clearpage
\end{document}